\pdfoutput=1 

\documentclass[]{llncs}

\usepackage[utf8]{inputenc}
\usepackage{amsmath,amssymb,amsfonts,mathrsfs,stmaryrd,wasysym}

\usepackage[usenames,svgnames,table]{xcolor}
\usepackage{multirow}

\usepackage{url}

\usepackage{paralist}

\usepackage{setspace}
\usepackage{calc}

\usepackage{tikz}
\usetikzlibrary{calc,matrix,positioning,fit,arrows,automata,shapes.misc,shapes.geometric,shapes.arrows,shapes.callouts,shapes.symbols,backgrounds,decorations,decorations.pathmorphing}

\usepackage{groove2tikz}

\usepackage{graphicx}

\usepackage{rotating}
\usepackage{adjustbox}

\usepackage{textcomp}

\usepackage{listings}

\usepackage{xspace}

\lstdefinestyle{basestyle}
{%
	frame=none,
	tabsize=2,
	numbers=none,
	basicstyle=\footnotesize\ttfamily,
	upquote=true,
	columns=fixed,
	showstringspaces=false,
	numberstyle=\ttfamily\tiny,
	numbersep=5pt,
	extendedchars=true,
	breaklines=true,
	showtabs=false,
	showspaces=false,
	showstringspaces=false,
	identifierstyle=\ttfamily,
	keywordstyle=\ttfamily\bfseries\color[rgb]{0,0,0.6},
	keywordstyle=[2]\ttfamily\bfseries\color{DarkRed},
	commentstyle=\it\ttfamily\color[rgb]{0,0.4,0},
	stringstyle=\it\ttfamily\color[rgb]{0.4,0,0},
	captionpos=b,
  numberbychapter=false,
	escapeinside={(*@}{@*)}
}
\lstdefinestyle{eiffellistingstyle}
{%
	style = basestyle,
	basicstyle=\scriptsize\ttfamily,
	language = Eiffel,
	morekeywords = {across, alias, all, and, as, check, class, interface, creation, create, debug, deferred, do, else, elseif, end, ensure, expanded, export, external, False, feature, from, frozen, if, implies, indexing, infix, inherit, inspect, invariant, is, like, local, loop, not, obsolete, old, once, or, prefix, redefine, rename, require, rescue, retry, select, some, strip, then, True, undefine, unique, until, variant, when, xor, Result, Current, Void, attached, detachable, agent, separate},
	morecomment=[l]--
}

\lstdefinestyle{eiffelinlinestyle}
{%
	style=eiffellistingstyle,
	basicstyle=\footnotesize\ttfamily
}

\lstnewenvironment{eiffelcode}[1][]
{%
	\lstset{style=eiffellistingstyle,#1}
}{}

\lstdefinestyle{controlprgstyle}
{%
	style = basestyle,
	basicstyle=\tiny\ttfamily,
	language = C,
	morekeywords = {alap, while, try},
	morekeywords=[2]{recipe},
}

\newcommand{\incode}[1]{\lstinline[style=eiffelinlinestyle]{#1}}

\AtBeginDocument{}

\usepackage[normalem]{ulem}

\usepackage{booktabs}

\usepackage[text size=tiny, backgroundcolor=blue!30,bordercolor=blue,linecolor=blue]{todonotes}
\makeatletter
\renewcommand{\todo}[2][]{\tikzexternaldisable\@todo[#1]{#2}\tikzexternalenable}
\makeatother

\newcommand{\myparagraph}[1]{\noindent\textbf{#1}\hspace{1.5ex}}

\newcommand{\gts}{\textsc{Gts}\xspace}
\newcommand{\scoop}{\textsc{Scoop}\xspace}
\newcommand{\groove}{\textsc{Groove}\xspace}

\newcommand{\scoopgts}{\scoop-\gts}
\newcommand{\scoopgraph}{\scoop-Graph\xspace}
\newcommand{\scoopgraphs}{\scoop-Graphs\xspace}
\newcommand{\qoq}{\textsc{QoQ}\xspace}
\newcommand{\req}{\textsc{RQ}\xspace}

\newcommand{\fifo}{\textsc{Fifo}\xspace}

\newcommand{\model}[1]{%
\ifx&#1& \scoopgts
\else \mbox{\scoopgts}(#1)\xspace
\fi}

\newcommand{\executionmodel}[1]{#1 execution model\xspace}
\newcommand{\qoqem}{\executionmodel{\qoq}}
\newcommand{\rqem}{\executionmodel{\req}}

\newcommand{\eve}{\textsc{Eve}\xspace}
\newcommand{\eveide}{\textsc{Eve} IDE\xspace}
\newcommand{\cfg}{\textsc{Cfg}\xspace}

\newcommand{\KK}{\ensuremath{\mathbb{K}}\xspace}

\newcommand{\maude}{\textsl{Maude}\xspace}

\newcommand{\eg}{e.g.\ }
\newcommand{\ie}{i.e.\ }

\newcommand*\samethanks[1][\value{footnote}]{\footnotemark[#1]}

\begin{document}
\title{A Graph-Based Semantics Workbench for\\ Concurrent Asynchronous Programs}
\titlerunning{A Graph-Based Semantics Workbench}


\author{Claudio Corrodi\inst{1,2}\thanks{
Research done whilst employed by the Chair of Software Engineering, \mbox{ETH Z\"{u}rich}.
} \and Alexander Heu{\ss}ner\inst{3} \and Christopher M. Poskitt\inst{1,4}\samethanks}
\authorrunning{C. Corrodi, A. Heu{\ss}ner, and C.M. Poskitt}

\institute{ Department of Computer Science, ETH Z\"{u}rich, Switzerland \and%
  Software Composition Group, University of Bern, Switzerland \and%
  Software Technologies Research Group, University of Bamberg, Germany
  \and%
  Singapore University of Technology and Design, Singapore }

\maketitle

\begin{abstract}
  A number of novel programming languages and libraries have been proposed that offer simpler-to-use models of concurrency than\linebreak[4] threads. It is challenging, however, to devise execution models that successfully realise their abstractions without forfeiting performance or introducing unintended behaviours. This is exemplified by \scoop---a concurrent object-oriented message-passing language---which has seen multiple semantics proposed and implemented over its evolution. We propose a  ``semantics workbench'' with fully and semi-automatic tools for \scoop, that can be used to analyse and compare programs with respect to different execution models. We demonstrate its use in checking the consistency of semantics by applying it to a set of representative programs, and highlighting a deadlock-related discrepancy between the principal execution models of the language. Our workbench is based on a modular and parameterisable graph transformation semantics implemented in the \groove tool. We discuss how graph transformations are leveraged to atomically model intricate language abstractions, and how the visual yet algebraic nature of the model can be used to ascertain soundness.
\end{abstract}

\section{Introduction}
To harness the power of multi-core and distributed architectures, software engineers must program with concurrency, asynchronicity, and parallelism in mind. Classical thread-based approaches to concurrent programming, however, are difficult to master and error prone. To address this, a number of programming APIs, libraries, and languages have been proposed that provide safer and simpler-to-use models of concurrency, such as the block-dispatch model of Grand Central Dispatch~\cite{GCD-Reference}, or the message-passing-based model of \scoop~\cite{West-NM15b}.
The concurrent programming abstractions that these languages provide rely on the existence of effective execution models for realising them; \emph{effective} in the sense that they do so without forfeiting performance or introducing unintended behaviours. Devising execution models that successfully reconcile these requirements, however, is challenging: a model that is too restrictive can deny desirable concurrency and lead to unnecessary bottlenecks; a model that is too permissive can lead to surprising and unexpected executions.

This challenge is exemplified by \scoop~\cite{West-NM15b}, a message-passing paradigm for concurrent object-oriented programming that aims to preserve the well-understood modes of reasoning enjoyed by sequential programs, such as pre- and postcondition reasoning over blocks of code. Although the high-level language mechanisms for achieving this were described informally as early as the `90s~\cite{Meyer93a,Meyer97a}, it took several years to understand how to effectively implement them: execution models~\cite{Brooke-PJ07a,Morandi-SNM13a,West-NM15b}, prototypes~\cite{Nienaltowski07a,Torshizi-OPC09a}, and contrasting versions of a production-level implementation~\cite{SCOOP-EiffelStudio-Reference} gradually appeared over the last decade, and can be seen as representing multiple, partially conflicting semantics for realising \scoop. They are also unlikely to be the last, as new language features continue to be proposed, prototyped, and integrated, \eg\cite{Morandi-NM14a}.
Despite the possible ramifications to behavioural and safety properties of existing programs, little work has been done to support formal and automatic \emph{comparisons} of the program executions permitted by these different semantics. While general, tool-supported formalisations exist---in \maude's conditional rewriting logic \cite{Morandi-SNM13a}, for example, and in a custom-built \textsc{Csp} model checker \cite{Brooke-PJ07a}---these are tied to particular execution models, do not operate on program source code, and are geared towards ``testing'' the semantics rather than general verification tasks. Furthermore, owing to the need to handle waiting queues, locks, asynchronous remote calls, and several other intricate features of the \scoop execution models, these formalisations quickly become complex, making it challenging to ascertain their soundness with language designers who lack a formal methods background.

\myparagraph{The Challenge.}
There is a need for languages like \scoop to have tools that not only support the prototyping of new semantics (and semantic extensions), but that also facilitate formal, automatic, and practical analyses for comparing the executions permitted by these semantics, and highlighting where behavioural and safety-related discrepancies arise. The underlying formalism for modelling the semantics should not be ad hoc; rather, it should support re-use, a modular design, and be easily extensible for language evolutions and changes. Furthermore, such tools should be usable in practice: the modelling formalism must be accessible to and understandable by software engineers, and the analyses must support several idiomatic uses of the language mechanisms.

\myparagraph{Our Contributions.}
We propose a ``semantics workbench'' equipped with fully and semi-automatic tools for \scoop, that can be used to analyse and compare programs with respect to different execution models for the purpose of checking their consistency. We demonstrate its use by formalising the two principal execution models of \scoop, analysing a representative set of programs with respect to both, and highlighting some behavioural and deadlock-related discrepancies that the workbench uncovers automatically. Our workbench is based on a modular and parameterisable graph transformation system (\gts) semantics, built upon our preliminary modelling ideas in \cite{GaM2015}, and implemented in the general-purpose \gts analysis tool \groove \cite{Ghamarian-MRZZ12a}. We leverage this powerful formalism to atomically model complex programmer-level abstractions, and show how its inherently visual yet algebraic nature can be used to ascertain soundness.
For language designers, this paper presents a transferable approach to checking the consistency of competing semantics for realising concurrency abstractions. For the graph transformation community, it presents our experiences of applying a state-of-the-art \gts tool to a non-trivial and practical problem in programming language design. For the broader verification community, it highlights a need for semantics-parameterised verification, and shows how \gts-based formalisms and tools can be used to derive an effective and modular solution. For software engineers, it provides a powerful workbench for crystallising their mental models of \scoop, thus helping them to write better quality code and (where need be) port it across different \scoop implementations.

\myparagraph{Plan of the Paper.}
After introducing the \scoop concurrency paradigm and its two most established execution models (Section~\ref{sec:scoop}), we introduce our formal modelling framework based on \gts, and show how to formalise different, parameterisable \scoop semantics (Section~\ref{sec:genericmodel}). Implementing our ideas in a small toolchain (Section~\ref{sec:tool}) allows us to check the consistency of semantics across a set of representative \scoop programs (Section~\ref{sec:analysis}), and highlight both a behavioural and deadlock-related discrepancy. To conclude, we summarise some related work (Section~\ref{sec:outro}), our contributions, and some future research directions (Section~\ref{sec:conclusion}).

\section{SCOOP and its Execution Models}\label{sec:scoop} 

\scoop~\cite{West-NM15b} is a message-passing paradigm for concurrent object-oriented programming that aims to preserve the well-understood modes of reasoning enjoyed by sequential programs; in particular, pre- and postcondition reasoning over blocks of code. This section introduces the programmer-level language mechanisms and reasoning guarantees of \scoop, as well as its two most established execution models. These will be described in the context of \scoop's production-level implementation for Eiffel~\cite{SCOOP-EiffelStudio-Reference}, but the ideas generalise to any object-oriented language (as explored, \eg for Java~\cite{Torshizi-OPC09a}).

\myparagraph{Language Mechanisms.}
In \scoop, every object is associated with a \emph{handler} (also called a \emph{processor}), a concurrent thread of execution with the exclusive right to call methods on the objects it handles. In this context, object references may point to objects with the same handler (\emph{non-separate} objects) or to objects with distinct handlers (\emph{separate} objects). Method calls on non-separate objects are executed immediately by the shared handler. To make a call on a separate object, however, a \emph{request} must be sent to the handler of that object to process it: if the method is a \emph{command} (\ie it does not return a result) then it is executed asynchronously, leading to concurrency; if it is a \emph{query} (\ie a result is returned and must be waited for) then it is executed synchronously. Note that handlers cannot synchronise via shared memory: only by exchanging requests.

The possibility for objects to have different handlers is captured in the type system by the keyword \incode{separate}. To request method calls on objects of \incode{separate} type, programmers simply make the calls within \emph{separate blocks}. These can be explicit (we will use the syntax \incode{separate x,y,}\ \dots\ \incode{do}\ \dots\ \incode{end}); but they also exist implicitly for methods with separate objects as parameters.

\myparagraph{Reasoning Guarantees.}
\scoop provides certain guarantees about the order in which calls in separate blocks are executed to help programmers avoid concurrency errors. In particular, method calls on separate objects will be logged as requests by their handlers in the order that they are given in the program text; furthermore, there will be no intervening requests logged from other handlers. These guarantees exclude object-level data races by construction, and allow programmers to apply pre- and postcondition reasoning within separate blocks independently of the rest of the program. Consider the following example (adapted from \cite{West-NM15b}), in which two distinct handlers are respectively executing blocks that set the ``colours'' of two \incode{separate} objects:\\[-2.5ex]
\noindent\begin{minipage}[t]{.45\textwidth}
	\vfill
 \begin{eiffelcode}
separate x,y
do
	x.set_colour (Green)
	y.set_colour (Green)
end
\end{eiffelcode}
\end{minipage}
\hfill
\begin{minipage}[t]{0.45\textwidth}
\begin{eiffelcode}
separate x,y
do
	x.set_colour (Indigo)
	a_colour = x.get_colour
	y.set_colour (a_colour)
end
\end{eiffelcode}
\end{minipage}

\noindent The guarantees ensure that whilst a handler is inside its \incode{separate x,y} block, the other handler cannot log intervening calls on \incode{x} or \incode{y}. Consequently, if the colours are later queried in another \incode{separate x,y} block, both of them will be Green or both of them will be Indigo; interleavings permitting any other combination to be observed are entirely excluded. This additional control over the order in which requests are processed represents a twist on classical message-passing models, such as the actor model~\cite{Agha86a}, and programming languages like Erlang~\cite{Armstrong-VW96a} that implement them.

\myparagraph{Execution Models.}
The abstractions of \scoop require an execution model that can realise them without forfeiting performance or introducing unintended behaviours. Two contrasting models have been supported by different versions of the implementation: initially, a model we call Request Queues (\req)~\cite{Morandi-SNM13a}, and a model that has now replaced it which we will call Queues of Queues (\qoq)~\cite{West-NM15b}.

The \rqem associates each handler with a single \fifo queue for storing incoming requests. To ensure the reasoning guarantees, each queue is protected by a lock, which another handler must acquire to be able to log a request on the queue. Realising a \incode{separate x,y,}\dots\ block then boils down to acquiring locks on the request queues attached to the handlers of \incode{x,y,}\dots\ and exclusively holding them for the duration of the block. This coarse-grained solution successfully prevents intervening requests from being logged, but leads to performance bottlenecks in several situations (\eg multiple handlers vying for the lock of a highly contested request queue).

In contrast, the \qoqem associates each handler with a \fifo queue that itself contains (possibly several) \fifo subqueues for storing incoming requests. These subqueues represent ``private areas'' for handlers to log requests without interference from other handlers. Realising a \incode{separate x,y,}\dots\ block no longer requires vying for locks; instead, the handlers of \incode{x,y,}\dots\ simply generate private subqueues on which requests can be logged without interruption for the duration of the block. If another handler also wants to log requests, then a new private subqueue is generated for it and its requests can be logged at the same time. The \qoq model removes the performance bottlenecks caused by the locks of \req, while still ensuring the \scoop reasoning guarantees by completely processing subqueues in the order that they were generated.

\begin{figure}[t!]
	\centering
	\includegraphics[width=0.5\textwidth]{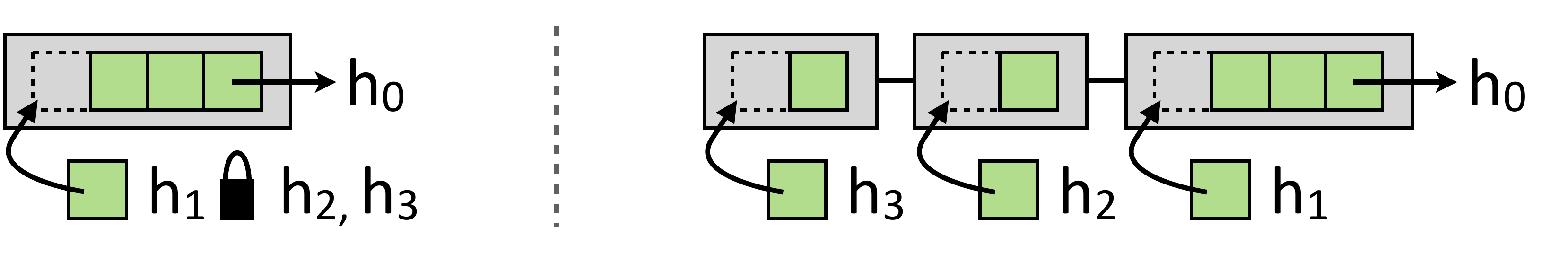}
  \vspace{-2ex}
	\caption{Logging requests under the \req (left) and \qoq (right) execution models}\label{fig:rq-qoq-graphics}
\end{figure}

Figure \ref{fig:rq-qoq-graphics} visualises three handlers ($h_1,h_2,h_3$) logging requests (green blocks) on another handler ($h_0$) under the two execution models. Note that the \req and \qoq implementations (i.e.~compilers and runtimes) include additional optimisations, and strictly speaking, can themselves be viewed as competing semantics.

\myparagraph{Semantic Discrepancies.}
Discrepancies between the execution models can arise in practice. In the mental model model of programmers, with \req, separate blocks had become synonymous with acquiring and holding locks---which are not implied by the basic reasoning guarantees or the \qoq model. This discrepancy comes to light with the classical dining philosophers program (as provided in the official \scoop documentation \cite{SCOOP-EiffelStudio-Reference}), which will form a running example for this paper. Under \req, Listing~\ref{listing:scoopdp2} (``eager'' philosophers) solves the problem by relying on the implicit parallel acquisition of locks on the forks' handlers; no two adjacent philosophers can be in their separate blocks (representing ``eating'') at the same time. Under \req, Listing~\ref{listing:scoopdp1} (``lazy'' philosophers) can lead to circular deadlocks, as philosophers acquire the locks in turn. With \qoq however---where there is no implicit locking---neither version represents a solution, and neither can cause a deadlock; yet the basic guarantees about the order of logged requests remain satisfied. We will return to this example in later sections, and show how such discrepancies can be detected by our workbench.

\begin{figure}[!bt]
  \centering
  \begin{minipage}[t]{0.4\textwidth}
  \begin{eiffelcode}[caption={Eager philosophers},label=listing:scoopdp2]
separate left_fork, right_fork
do
  left_fork.use
  right_fork.use
end
(*@\vspace{5ex}@*)
  \end{eiffelcode}
  \end{minipage}
  \hspace{0.5cm}
  \begin{minipage}[t]{0.4\textwidth}
    \begin{eiffelcode}[caption={Lazy philosophers},label=listing:scoopdp1]
separate left_fork
do
  separate right_fork
  do
    left_fork.use
    right_fork.use
  end
end
    \end{eiffelcode}
  \end{minipage}
  \vspace{-2ex}
\end{figure}


\section{A Graph-based Semantic Model for the SCOOP Family}\label{sec:genericmodel}

There are several established and contrasting semantics of \scoop \cite{Brooke-PJ07a,GaM2015,Morandi-SNM13a,Ostroff-THS09a,West-NM15b} including a comprehensive reference semantics for \req in \maude{}'s conditional rewriting logic \cite{Morandi-SNM13a}, and a semantics for the core of \qoq in the form of simple structural operational rules \cite{West-NM15b}. These formalisations, however, cannot easily be used for semantic comparisons, due to their varying levels of detail, coverage, extensibility, and tool support. Hence we present in this section ``yet another'' semantic model, called \scoopgts, based on our preliminary modelling ideas for \req in~\cite{GaM2015}, using the formalism of graph transformation systems (\gts).

Our reasons to introduce \scoopgts are manifold:
\begin{inparaenum}[(a)]
  \item we need a common modelling ground that can be parameterised by models of \req and \qoq;
  \item known models based on algebra, process calculi, automata, or Petri nets do not straightforwardly cover \scoop's asynchronous concurrent nature, or would hide these features in intricate encodings; 
  \item existing approaches are often proposed from a theoretician's point of view and are not easily readable by software engineers, whereas graphs and diagrammatic notations (\eg \textsc{Uml}) might already be used in their everyday work.
\end{inparaenum}
Choosing graph transformations as our base formalism is well-justified, as they satisfy the above requirements, and reconcile the goal to have a theoretically rigorous formalisation with the goal to be accessible to software engineers, \eg for expert interviews with the language implementers (see~\cite{Rensink10} for a detailed discussion of the pros and cons of \gts in this setting). The ``non-linear'' context of graph rewriting rules proves to be a powerful mechanism for defining semantics and their interfaces for parameterisation.

We formalised \scoopgts using the state-of-the-art \gts tool \groove~\cite{GROOVE-Reference}. Due to limited space, we provide all the files necessary to browse our \gts model as supplementary material \cite{supplement}, including input graphs generated from the example programs of Section~\ref{sec:analysis} that can be simulated, analysed, and verified.

\begin{figure}[t]
	\centering
	\includegraphics{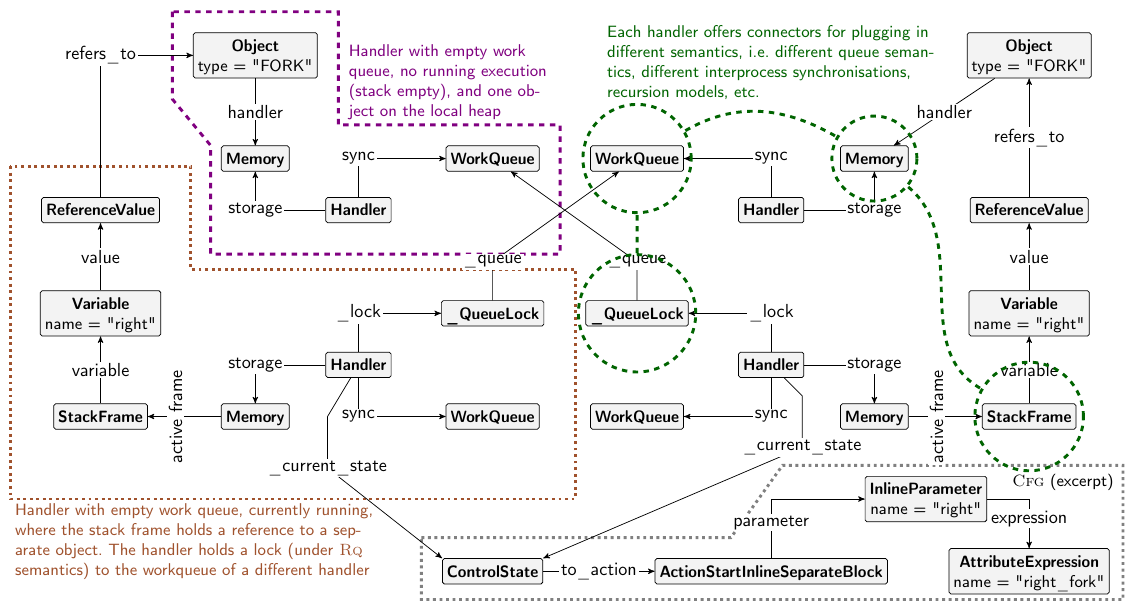}
  \vspace{-2ex}
  \caption{Reachable deadlock under \req for the lazy philosophers program (Listing~\ref{listing:scoopdp1}) simplified from \groove output   with additional highlighting and information in colour}
	\label{fig:deadlock_rq}
\end{figure}

\myparagraph{SCOOP-Graphs.}
Each global configuration of a \scoop program, \ie snapshot of the global state, is represented by a directed, typed attributed graph consisting of
\begin{inparaenum}[(i)]
  \item \emph{handler nodes} representing \scoop's handlers, \ie basic execution units;
  \item a representation of each handler's \emph{local memory} (\ie ``heap'' of non-separate objects) and its known neighbourhood, consisting of references to separate objects that can be addressed by queries and commands;
  \item a representation of each handler's \emph{stack}, via stack frames that model recursive calls to non-separate objects;
  \item \emph{requests} for modelling separate calls, which are stored in
  \item subgraphs representing each handler's \emph{input work queue};
  \item a global \emph{control flow graph} (\cfg)  presenting the program's execution blocks (consisting of states and actions/transitions in-between);
  \item  relations to model inter-handler and handler-memory relations (\eg locking, waiting, etc.) and to assign each handler to its current state in the \cfg; and
  \item additional bookkeeping nodes, \eg containing information on detected deadlocks, and nodes to model the interfaces/contexts for semantic parameterisation.
\end{inparaenum}
An example \scoopgraph can be seen in Figure~\ref{fig:deadlock_rq}, depicting a configuration with two concurrently running and two idle handlers.

\begin{figure}[tb]
  \centering
	\includegraphics{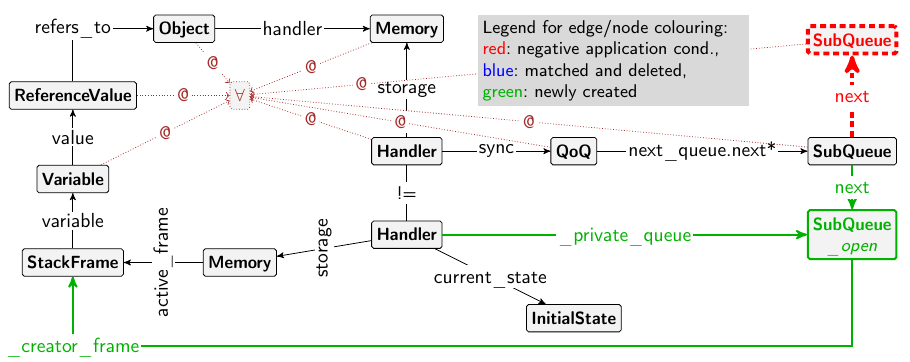}
  \vspace{-2.5ex}
  \caption{Simplified \qoq rule for entering a \incode{separate x,y,}\dots\ block, which uses $\forall$-quantification to atomically match arbitrarily many handlers. The rule assumes that the handlers' queues already contain some other private subqueues open}\label{fig:reserve_handlers_qoq}
\end{figure}

\myparagraph{GTS-based Operational Semantics.}
The operational semantics of \scoopgts is given by graph-rewriting rules that are regimented by \emph{control programs}.
An example rule, concisely written using nesting as supported by \groove, can be seen in Figure~\ref{fig:reserve_handlers_qoq}. Note that nested rules (including $\forall$- and $\exists$-quantification) allow us to express complex, atomic rule matchings in a relatively straightforward and brief way (compared to rules in classical operational semantics, \eg in \cite{West-NM15b} for multiple handler reservations).
A simplified, example control program can be seen in Listing~\ref{listing:controlprg}.  Control programs allow us to make an execution model's scheduler explicit (and thus open to parameterisation) and help us to implement ``garbage collection'' for the model (\eg removing bookkeeping edges no longer needed). Furthermore, they provide a fine-grained way to control the atomicity of \scoop operations, by wrapping sequences of rule applications into so-called \emph{recipes}.

\myparagraph{Semantic Modularity of \scoopgts.}
We support semantic parameterisation for \scoopgts by providing fixed module interfaces in the graph via special ``plug-in nodes/edges'' (\eg \incode{WorkQueue}, \incode{Memory}, \incode{StackFrame} in Figure~\ref{fig:deadlock_rq}), and changing only the set of \gts rules that operate on the subgraphs that they guard. We have modelled both \req and \qoq with distinct sets of rules that operate on the subgraphs guarded by \incode{WorkQueue}: we call the model parameterised by \req and \qoq respectively \model{\req} and \model{\qoq}. As well as parameterising the queue semantics, it is possible to model different recursion schemes, memory models, and global interprocess synchronisations.

This semantic modularity also permits us to directly apply abstractions to \scoopgts, \eg changing the queue's semantics to a bag's counting abstraction, or flattening recursion. This could prove useful for providing advanced verification approaches in the workbench.

\begin{lstlisting}[float=tp,style=controlprgstyle,label=listing:controlprg,caption={Simplified control program (in \groove syntax)}]
initialize_model;                                 // call gts rule for initialisation
while (progress & no_error) {
  for each handler p:                             // choose handlers under some scheduling strategy
    alap handler_local_execution_step(p)+;        // each handler executes local actions as long as possible
  try synchronisation_step;                       // then try (one) possible global synchronisation step
}
recipe handler_local_execution_step (p){
  try separate_object_creation(p)+;               // try local actions that are possibly applicable
  else try assignment_to_variable(p)+;
  else try ... ;                                  // sequentially try all other possible actions
  try clean_up_model+;                            // do some "garbage collection" to keep the model small
}
recipe synchronisation_step(){
  reserve_handlers | dequeue_task | ...;          // non-deterministically try to synchronise
}
...                                               // remaining recipes (core functionality)
// ---------- plug in -------------------------------------------------------------------------------
recipe separate_object_creation(p){               // provide different implementations for RQ and QoQ
  ...                                             // and parameterise the control program
}
...                                               // remaining recipes that are plugged in
\end{lstlisting}

\begin{figure}[!t]
\includegraphics[scale=0.95]{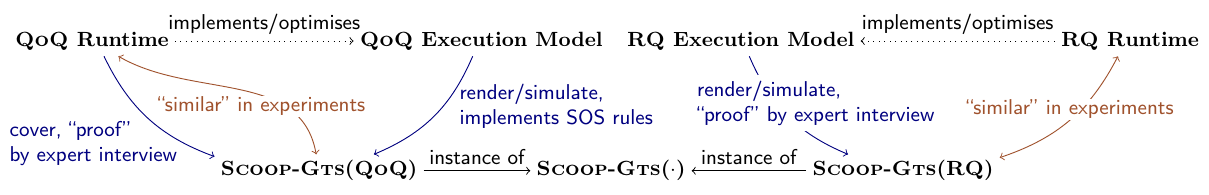}
\vspace{-2.5ex}
  \caption{Relation between \scoopgts, the execution models, and the runtimes}\label{fig:soundness}
\end{figure}

\myparagraph{Soundness/Faithfulness.}
The relation of \scoopgts to the most prominent execution models and runtimes is depicted in Figure~\ref{fig:soundness}. Due to the varying levels of detail in the formalisations of the \req and \qoq execution models (and lack of formalisations of their implementations/runtimes), there is no universal way to prove \scoopgts's faithfulness to them. We also remark that \scoopgts currently does not support some programming mechanisms of the Eiffel language (e.g.~exceptions, agents), but could be straightforwardly extended to cover them.

We were able to conduct expert interviews with the researchers proposing the execution models and the programmers implementing the \scoop compiler and runtimes, which helped to improve our confidence that \scoopgts faithfully covers their behaviour. Here, \scoopgts's advantage of a visually accessible notation was extremely beneficial, as we were able to directly use simulations in \groove during the interviews, which were understood and accepted by the interviewees.
In addition, we compared \groove simulations of the executions of \scoop programs (see the benchmarks of Section~\ref{sec:analysis}) against their actual execution behaviour in the official \scoop IDE and compiler (both the current release that implements \qoq, and an older one that implemented \req). Again, this augmented our confidence.
Furthermore, we were able to compare \model{\qoq} with the structural operational semantics for \qoq provided in \cite{West-NM15b}. Unfortunately, the provided semantic rules focus only on a much simplified core, preventing a rigorous bisimulation proof exploiting the algebraic characterisations of \gts. We can, however, straightforwardly implement and simulate them in our model.

To conclude, \scoopgts fits into the suite of existing \scoop formalisations, and is able to cover (avoiding the semantically overloaded word ``simulate'') both of the principal execution models.

\myparagraph{Expressiveness.}
As previously discussed, \scoopgts is expressive enough to cover the existing \req and \qoq semantic models of \scoop due to its modularity and the possibility to plug-in different queueing semantics.
We currently plan to include other semantic formalisations of \scoop-like languages, \eg the concurrent Eiffel proposed by~\cite{Brooke-Paige09a} (similar to \scoop but differences regarding separate object calls), other actor-based object-oriented languages, and concurrency concepts like ``co-boxes''~\cite{schaefer2010}.
\scoopgts is obviously Turing-complete (one can simulate a 2-counter Minsky machine by non-recursive models with one object per handler, similar to the construction in~\cite{Geeraerts-HR13a}).  A proper formal investigation into its computational power (also that of subclasses of the model) is ongoing.

\section{Toolchain for the Workbench}\label{sec:tool}

\newcommand{\eselsohr}[1]{%
    \draw[white] (#1) -- +(-0.2,0) -- +(0,-0.2) -- cycle;
    \draw[black] (#1)  +(-0.2,0) -- +(-0.2,-0.2) -- +(0,-0.2) -- cycle;
}

\begin{figure}[!t]
	\centering
	\includegraphics[scale=0.95]{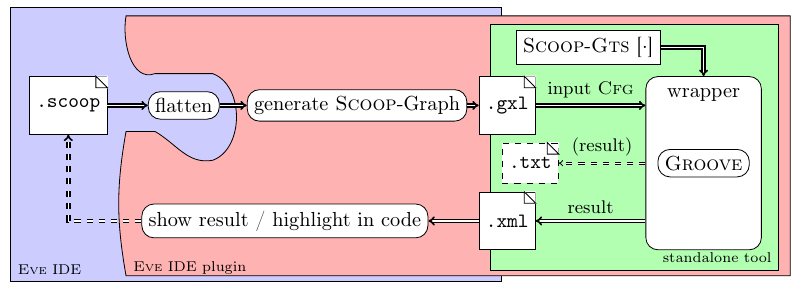}
  \vspace{-2ex}
  \caption{Overview of our toolchain: a plugin integrated with the (research version of the) official \scoop IDE, which interfaces with a wrapper that utilises and controls \groove in the background. The wrapper can also be used as part of a standalone tool}
	\label{fig:toolchain}
\end{figure}
Our semantics workbench consists of a toolchain that bridges the gap between \scoop program code and the analysis of \scoopgts models in \groove. In particular, it translates source code into \scoopgraph{}s, executes the appropriate analyses in \groove, and then collects and returns the results to the user.

\begin{figure}[tb]
  \centering
	\begin{adjustbox}{width=\textwidth,max height=\textheight}
%
\begin{tikzpicture}[scale=\tikzscale]
\node[basic_node] (n7) at (2.810, -7.550) {\ml{\textbf{InitialState}\\class = "PHILOSOPHER"\\procedure = "lazy\_eat"}};
\node[basic_node] (n15) at (5.930, -7.640) {\ml{\textbf{ActionStartInlineSeparateBlock}}};
\node[basic_node] (n16) at (5.994, -6.648) {\ml{\textbf{InlineParameter}\\name = "left"}};
\node[basic_node] (n17) at (8, -6.648) {\ml{\textbf{AttributeExpression}\\name = "left\_fork"}};
\node[basic_node] (n18) at (9.160, -7.700) {\ml{\textbf{ControlState}}};
\node[basic_node] (n21) at (12.230, -7.750) {\ml{\textbf{ActionStartInlineSeparateBlock}}};
\node[basic_node] (n20) at (12.240, -6.580) {\ml{\textbf{InlineParameter}\\name = "right"}};
\node[basic_node] (n22) at (10, -6.580) {\ml{\textbf{AttributeExpression}\\name = "right\_fork"}};

\begin{scope}[yshift=16]
  \node[basic_node] (n23) at (2.720, -10.230) {\ml{\textbf{ControlState}}};
  \node[basic_node] (n25) at (4.970, -10.180) {\ml{\textbf{ActionCommand}\\procedure = "use"}};
  \node[basic_node] (n26) at (4.960, -9.3) {\ml{\textbf{LocalExpression}\\name = "left"}};
  \node[basic_node] (n27) at (7.050, -10.200) {\ml{\textbf{ControlState}}};
  \node[basic_node] (n29) at (9.550, -10.210) {\ml{\textbf{ActionCommand}\\procedure = "use"}};
  \node[basic_node] (n28) at (9.534, -9.3) {\ml{\textbf{LocalExpression}\\name = "right"}};
  \node[basic_node] (n30) at (11.840, -10.180) {\ml{\textbf{ControlState}}};
\end{scope}

\begin{scope}[yshift=22]
\node[basic_node] (n19) at (6.354, -11.434) {\ml{\textbf{ActionEndInlineSeparateBlock}}};
  \node[basic_node] (n24) at (11.808, -11.360) {\ml{\textbf{ActionEndInlineSeparateBlock}}};
  \node[basic_node] (n31) at (9.054, -11.398) {\ml{\textbf{ControlState}}};
  \node[basic_node] (n32) at (2.632, -11.482) {\ml{\textbf{FinalState}}};
\end{scope}

\path[basic_edge](n7.east |- 5.930, -7.640) -- node[lab] {\ml{to\_action}} (n15) ;
\path[basic_edge](n16) -- node[lab,rotate=90] {\ml{expression}} (n17) ;
\path[basic_edge](n15.north -| 5.994, -6.648) -- node[lab] {\ml{parameter}} (n16) ;
\path[basic_edge](n15.east |- 9.160, -7.700) -- node[lab] {\ml{to\_state}} (n18) ;
\path[basic_edge] (n15.south)  -- ++(0,-0.2) -- ++(-4,0) -- node[lab]{\ml{\_ends\_with}}++(0,-2) -| (n19);
\path[basic_edge](n20) -- node[lab,rotate=90] {\ml{expression}} (n22) ;
\path[basic_edge](n21.north -| 12.240, -6.580) -- node[lab] {\ml{parameter}} (n20) ;
\path[basic_edge] (n21)  -- ++(0,-0.45) -| node[lab] {\ml{to\_state}}  (n23);
\path[basic_edge](n18.east |- 12.230, -7.750) -- node[lab] {\ml{to\_action}} (n21) ;
\path[basic_edge](n21.south -| 12.940, -10.210) -- (12.940, -10.210) --  (n24)
node[lab] at (12.940, -9.747) {\ml{\_ends\_with}};
\path[basic_edge](n25) -- node[lab] {\ml{to\_state}} (n27) ;
\path[basic_edge](n25) -- node[lab] {\ml{target}} (n26) ;
\path[basic_edge](n27) -- node[lab] {\ml{to\_action}} (n29) ;
\path[basic_edge](n29.north -| 9.534, -9.032) -- node[lab] {\ml{target}} (n28) ;
\path[basic_edge](n23) -- node[lab] {\ml{to\_action}} (n25) ;
\path[basic_edge](n29) -- node[lab] {\ml{to\_state}} (n30) ;
\path[basic_edge](n30.south -| 11.808, -11.360) -- node[lab] {\ml{to\_action}} (n24) ;
\path[basic_edge](n24) -- node[lab] {\ml{to\_state}} (n31) ;
\path[basic_edge](n31) -- node[lab] {\ml{to\_action}} (n19) ;
\path[basic_edge](n19) -- node[lab] {\ml{to\_state}} (n32) ;
\end{tikzpicture}
  \end{adjustbox}
  \vspace{-3ex}
	\caption{Generated control-flow graph for Listing~\ref{listing:scoopdp1}}
	\label{fig:cfg_lazy_eat}
\end{figure}
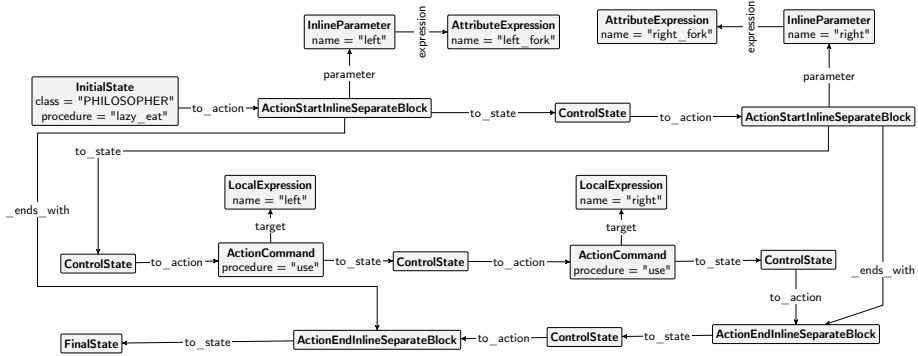

Our toolchain is summarised in Figure~\ref{fig:toolchain}. Its principal component is a plug-in for the \eveide---a research version of the \scoop/Eiffel IDE (including the production compiler and runtime) which supports the integration of verification tools \cite{Tschannen-FNM11a}. For a given \scoop program, the plug-in uses the existing services of \eve to check that the code compiles, and then extracts a representation of it in which inheritance has been ``flattened''. From this flattened program, we generate a \scoopgraph (encoded in the Graph eXchange Language) which corresponds very closely to the abstract syntax tree of the original program. See, for example, Figure~\ref{fig:cfg_lazy_eat}, which is generated from the code in Listing~\ref{listing:scoopdp1}. Observe that between the \incode{InitialState} and \incode{FinalState}, the control-flow graph directly encodes the four actions of the original program: two declarations of \incode{separate}blocks, and two commands within them.
We provide a wrapper (written in Java) around the external \groove tool, which takes a generated \scoopgraph as input, and launches a full state-space exploration in \groove with respect to  \model{\req} or \model{\qoq}. The results---including statistics and detected error states---are then extracted from \groove and returned to the programmer for inspection. A standalone version of this wrapper without the \eve integration is also available and can be downloaded from \cite{supplement}.

\myparagraph{Checking the Consistency of Semantics.}
The workbench can be used to check the consistency of program executions under \req and \qoq with respect to various properties. These properties are encoded in \scoopgts as \emph{error rules} that match on configurations if and only if they violate the properties. If they match, they generate a special \incode{Error} node that encodes some contextual information for the toolchain to extract, and prevents the execution branch from being explored any further. Two types of error rules are supported: general, safety-related error rules for detecting problems like deadlock (whether caused by waiting for request queue locks in \req, or waiting on cycles of queries in \qoq); but also user-specified error rules for program-specific properties (as we will use in Section \ref{sec:analysis}). If any of these error rules are applied in a state-space exploration, this information is extracted and reported by the workbench toolchain; discrepancies between semantics exist when such rules match under only one.
Figure~\ref{fig:deadlock_rq} shows an actual deadlock between two handlers attempting to enter the nested separate block of Listing~\ref{listing:scoopdp1} under \req. This configuration is matched by an error rule for deadlock (not shown), which catches the circular waiting dependencies exhibited by the edges.

\section{Evaluation}\label{sec:analysis}

To evaluate the use of our workbench for checking the consistency of semantics, we devised a representative set of benchmark programs, based on documented \scoop examples \cite{SCOOP-EiffelStudio-Reference} and classical synchronisation problems. We then deployed the toolchain to analyse their executions under \req and \qoq with respect to behavioural and safety-related properties, and highlight the discrepancies uncovered by the workbench for our running example. Everything necessary to reproduce our evaluation is available at~\cite{supplement}.

\myparagraph{Benchmark Selection.}
Our aim was to devise a set of representative programs covering different, idiomatic usages of \scoop's concurrency mechanisms. To achieve this, we based our programs on official, documented examples \cite{SCOOP-EiffelStudio-Reference}, as well as some classical synchronisation problems, in order to deploy the language mechanisms in a greater variety of usage contexts. Note that it is not (yet) our goal to analyse large software projects, but rather to compare executions of representative programs with manageable state spaces under different semantics.

We selected the following programs: dining philosophers---as presented in Section~\ref{sec:scoop}---with its two implementations for picking up forks (eagerly or lazily) which exploited the implicit locking of \req; a third variant of dining philosophers without any commands in the separate blocks; single-element producer consumer, which uses a mixture of commands, queries, and condition synchronisation; and finally, barbershop and dining savages (based on \cite{little-book-of-semaphores}), both of which use a similar mix of features. These programs cover different usages of \scoop's language mechanisms and are well-understood examples in concurrent programming. Note that while our compiler supports inheritance by flattening the used classes, these examples do not use inheritance; in particular, no methods from the implicitly inherited class \incode{ANY} are used. By not translating these methods into the start graphs, we obtain considerably smaller graphs (which impacts the exploration speed, but not the sizes of the generated transition systems).

\myparagraph{Benchmark Results.}
Table~\ref{tab:performance} contains metrics for the inspected examples, obtained using our \groove wrapper utility. The presented values correspond to full state-space exploration. Metrics for elapsed time (wall clock time) and memory usage (computed using Java's \texttt{MemoryPoolMXBean}) are the means of five runs, while the other values are the same for each run. The experiments were carried out on an off-the-shelf notebook with an Intel Core i7-4810MQ CPU and 16 GB of main memory. We used Oracle Java 1.8.0\_25 with the \verb|-Xmx 14g| option together with \groove 5.5.5.

\begin{table*}[!t]\centering
	\scriptsize
  \caption{Results for the dining philosophers (DP, with the number of philosophers), producer-consumer (PC, with the number of elements), barbershop (with the number of customers), and dining savages (with the number of savages) programs;  time and memory metrics are means over five runs (standard deviation in brackets)}
	\label{tab:performance}
  \vspace{-2ex}
  \renewcommand{\arraystretch}{1.2}
  \newcommand{\greyrow}{\rowcolor{gray!40}}
  \newcommand{\sd}[1]{\makebox[7ex][r]{[#1]}}
  \newcommand{\sdsmall}[1]{\makebox[6ex][r]{[#1]}}
	\begin{adjustbox}{max width=10cm}
    \begin{tabular}{l@{}c@{ }r@{  }r@{ }c@{ }c@{}r@{ }r}
    \toprule\rowcolor{white}
    Start graph & Semantics & Config- & \multicolumn{1}{c}{Rule}  & Start graph size & Final graph size & \multicolumn{1}{c}{Time [SD]} & \multicolumn{1}{c}{Memory [SD]}\\
    \rowcolor{white}& & ~urations  & ~~applications & (nodes/edges) & (nodes/edges) & \multicolumn{1}{c}{(seconds)} & \multicolumn{1}{c}{(GB)} \\
		\midrule
DP 2 eager & RQ & 4219 & 54441 & 226/343 & 261/396 & 19.34 \sd{0.25} & 1.60 \sdsmall{0.02}\\
& QoQ & 5644 & 72762 & 226/343 & 284/462 & 25.61 \sd{1.03} & 1.63 \sdsmall{0.02}\\
\greyrow DP 2 lazy & RQ & 5679 & 72692 & 221/334 & 288/470 & 24.76 \sd{0.34} & 2.05 \sdsmall{0.10}\\
\greyrow
& QoQ & 9609 & 123583 & 221/334 & 256/387 & 42.46 \sd{0.65} & 2.22 \sdsmall{0.06}\\

DP 2 eager  & RQ & 442 & 6010 & 254/395 & 300/473 & 6.08 \sd{0.28} & 0.57 \sdsmall{0.00}\\
~~(no commands) & QoQ & 443 & 6135 & 254/395 & 300/473 & 5.84 \sd{0.24} & 0.57 \sdsmall{0.01}\\
\greyrow
DP 2 lazy &  RQ & 868 & 11211 & 250/387 & 325/541 & 9.53 \sd{0.36} & 0.66 \sdsmall{0.00}\\
\greyrow
~~(no commands) & QoQ & 919 & 11935 & 250/387 & 296/465 & 10.66 \sd{0.61} & 0.66 \sdsmall{0.01}\\
DP 3 eager & RQ & 99198 & 1270216 & 226/343 & 277/422 & 469.94 \sd{11.93} & 11.14 \sdsmall{0.17}\\
 & QoQ & 199144 & 2532882 & 226/343 & 304/498 & 1393.95 \sd{31.06} & 13.88 \sdsmall{0.08}\\
\greyrow
DP 3 lazy & RQ & 170249 & 2166712 & 221/334 & 319/536 & 1149.07 \sd{53.59} & 13.33 \sdsmall{0.69}\\
\greyrow
 & QoQ & 444686 & 5683419 & 221/334 & 272/413 & 2564.45 \sd{55.94} & 11.99 \sdsmall{0.07}\\
DP 3 eager  & RQ & 3269 & 43967 & 254/395 & 316/499 & 37.31 \sd{1.68} & 1.70 \sdsmall{0.09}\\
~~(no commands)& QoQ & 3286 & 45152 & 254/395 & 316/499 & 39.32 \sd{0.65} & 1.52 \sdsmall{0.00}\\
\greyrow
DP 3 lazy & RQ & 10877 & 139216 & 250/387 & 355/604 & 114.26 \sd{5.17} & 3.39 \sdsmall{0.13}\\
\greyrow
~~(no commands) & QoQ & 11774 & 151526 & 250/387 & 312/491 & 125.25 \sd{4.19} & 3.63 \sdsmall{0.07}\\
PC 5 & RQ & 4085 & 51283 & 307/476 & 353/548 & 45.87 \sd{0.74} & 2.07 \sdsmall{0.14}\\
 & QoQ & 12366 & 156210 & 307/476 & 353/548 & 140.86 \sd{2.83} & 3.14 \sdsmall{0.07}\\
\greyrow
PC 20 & RQ & 12890 & 159958 & 307/476 & 398/593 & 148.28 \sd{2.85} & 3.78 \sdsmall{0.25}\\
\greyrow
 & QoQ & 50286 & 632820 & 307/476 & 398/593 & 618.33 \sd{15.42} & 6.67 \sdsmall{0.20}\\
Barbershop 2 & RQ & 38509 & 494491 & 302/466 & 346/538 & 354.17 \sd{16.58} & 6.91 \sdsmall{0.19}\\
 & QoQ & 54325 & 702611 & 302/466 & 346/538 & 537.21 \sd{8.19} & 7.79 \sdsmall{0.16}\\
\greyrow
Savages 2 & RQ & 35361 & 448576 & 410/631 & 459/716 & 550.92 \sd{14.99} & 7.06 \sdsmall{0.07}\\
\greyrow
 & QoQ & 79398 & 1008596 & 410/631 & 459/716 & 1299.52 \sd{54.96} & 11.33 \sdsmall{0.11}\\
		\bottomrule
	\end{tabular}
\end{adjustbox}
\end{table*}

Across all instances, the start and final graph sizes are comparable. This can be explained by the fact that our implementation contains a number of simple ``garbage collection'' rules that remove edges and nodes that are no longer needed (\eg the results of intermediate computations). Final graphs simply contain the control-flow graph and heap structure after the executions. Note that we do not perform real garbage collection. For example, unreachable objects are not removed; the graph size increases linearly with the number of created objects.

The number of configurations denotes the number of recipe applications. This value is of interest because it allows one to directly compare explorations under different semantics (\ie how much more concurrency is permitted). Recall that scheduler-specific rules are wrapped inside recipes. For example, enqueueing a work item may trigger more bookkeeping rules in \qoq than in \req. Since the corresponding logic (see Listing~\ref{listing:controlprg}) is implemented in a recipe, we end up with just one more configuration in both cases, independently of how many individual rule applications are triggered within the recipe. Differences in the number of configurations arise from different branching at synchronisation points. For example, we can see that in most instances, \qoq generates considerably more configurations than the \req implementation, which suggests that \scoop programs are ``more concurrent'' under \qoq.

The time and memory columns show the raw power requirements of our toolchain. Unfortunately, the state-space explosion problem is inevitable when exploring concurrent programs. The number of configurations is, unsurprisingly, particularly sensitive to programs with many handlers and only asynchronous commands (\eg dining philosophers). Programs that also use synchronous queries (\eg producer-consumer) scale better, since queries force synchronisation once they reach the front of the queue. We note again that our aim was to facilitate automatic analyses of representative \scoop programs that covered the different usages of the language mechanisms, rather than optimised verification techniques for production-level software. The results suggest that for this objective, the toolchain scales well enough to be practical.

\myparagraph{Error Rules / Discrepancies Detected.}
In our evaluation of the various dining philosophers implementations, we were able to detect that the lazy implementation (Listing~\ref{listing:scoopdp1}) can result in deadlock under the \req model, but not under \qoq.
This was achieved by using error rules that match circular waiting dependencies. In case a deadlock occurs that is not matched by these rules, we can still detect that the execution is stuck and report a generic error, after which we manually inspect the resulting configuration. While such error rules are useful for analysing \scoopgraphs in general, it is also useful to define rules that match when certain program-specific properties hold. For example, if we take a look at the eager implementation of the dining philosophers (Listing~\ref{listing:scoopdp2}) and its executions under \req and \qoq, we find that the program cannot deadlock under either. This does not prove however that the implementation actually solves the dining philosophers problem under both semantics. To check this, we defined an error rule that matches if and only if two adjacent philosophers are in their separate blocks at the same time, which is impossible if forks are treated as locks (as they implicitly are under \req). Consequently, this rule matches only under the \qoq semantics, highlighting that under the new semantics, the program is no longer a solution to the dining philosophers problem. (We remark that it can be ``ported'' to \qoq by replacing the commands on forks with queries, which force the waiting.) We implemented program-specific correctness rules for the other benchmark programs analogously, but did not detect any further discrepancies.


\section{Related Work}\label{sec:outro}
We briefly describe some related work closest to the overarching themes of our paper: frameworks for semantic analyses, \gts models for concurrent asynchronous programs, and verification techniques for \scoop.

\myparagraph{Frameworks for Semantic Analysis.}
The closest approach in spirit to ours is the work on \KK \cite{DBLP:conf/wrla/LucanuSR12,DBLP:journals/jlp/RosuS10}.
It consists of the \KK concurrent rewrite abstract machine and the \KK technique. One can think of \KK as domain specific language for implementing programming languages with a special focus on semantics, which was recently successfully applied to give elaborate semantics to Java~\cite{KJavaPOPL2015} and JavaScript~\cite{park-stefanescu-rosu-2015-pldi}.  Both \KK and our workbench have the same user group (programming language designers and researchers) and focus on formalising semantics and analysing programs based on this definition. We both have ``modularity'' as a principal goal in our agendas, but in a contrasting sense: our modularity targets a semantic plug-in mechanism for parameterising different model components, whereas \KK focuses on modularity with respect to language feature reuse. In contrast to our approach, \KK targets the whole language toolchain---including the possibility to define a language and automatically generate parsers and a runtime simulation for testing the formalisation. Based on \maude's formal power of conditional rewriting logic, \KK also offers axiomatic models for formal reasoning on programs and the possibility to also define complex static semantic features, \eg advanced typing and meta-programming. Despite having similar formal underlying theoretical power (\KK's rewriting is similar to ``jungle rewriting'' graph grammars~\cite{DBLP:conf/gg/SerbanutaR12}), \scoopgts models make the graph-like interdependencies between concurrently running threads or handlers a first-class element of the model. This is an advantage for analyses of concurrent asynchronous programs, as many concurrency properties can be straightforwardly reduced to graph properties (\eg deadlocks as wait-cycles). Our explicit \gts model also allows us to compare program executions under different semantics, which is not a targeted feature of \KK. We also conjecture that our diagrammatic notations are easier for software engineers to grasp than purely algebraic and axiomatic formalisations.

\myparagraph{\gts Models for Concurrent Asynchronous Programs.}
Formalising and analysing concurrent object-oriented programs with \gts-based models is an emerging trend in software specification and analysis, especially for approaches rooted in practice. See \cite{Rensink10} for a good overview discussion---based on a lot of personal experience---on the general appropriateness of \gts for this task.
In recent decades, conditional rewriting logic has become a reference formalism for concurrency models in general; we refer to~\cite{DBLP:journals/tcs/Meseguer92} and its recent update~\cite{DBLP:journals/jlp/Meseguer12} for details. Despite having a comparable expressive power, our approach's original decision for \gts and for \groove as our state-space exploration tool led us to an easily accessible, generic, and parameterisable semantic model and tools that work in acceptable time on our representative \scoop examples.
Closest to our \scoopgts model is the \textsc{Qdas} model presented in \cite{Geeraerts-HR13a}, which represents an asynchronous, concurrent waiting queue based model with \emph{global memory} as \gts, for verifying programs written in Grand Central Dispatch \cite{GCD-Reference}. Despite the formal work, there is currently no direct compiler to \gts yet.
 The Creol model of \cite{JohnsenOY06} focuses on asynchronous concurrent models but without more advanced remote calls via queues as needed for \scoop. Analysis of the model can be done via an implementation in \maude~\cite{JohnsenOA05}.
Existing \gts-based models for Java only translate the code to a typed graph similar to the control-flow sub-graph of \scoopgts \cite{CorradiniDFR04,RensinkZ09}. A different approach is taken by \cite{FerreiraFR07}, which abstracts a \gts-based model for concurrent OO systems~\cite{FerreiraR05}  to a finite state model that can be verified using the SPIN model checker.
\groove itself was already used for verifying concurrent distributed algorithms on an abstract \gts level~\cite{Ghamarian-MRZZ12a}, but not on an execution model level as in our approach. However, despite the intention to apply generic frameworks for the specification, analysis, and verification of object-oriented concurrent programs, \eg in \cite{DottiDFRRS05,ZambonR11}, there are no publicly available tools implementing this long-term goal that are powerful enough for \scoop.

\myparagraph{\scoop Analysis/Verification.}
Various analyses for \scoop programs have been proposed, including: using a \scoop virtual machine for checking temporal properties~\cite{Ostroff-THS09a}; checking Coffman's deadlock conditions using an abstract semantics~\cite{Caltais-Meyer14a};  and statically checking code annotated with locking orders for the absence of deadlock~\cite{West-NM10a}. In contrast to our work, these approaches are tied to particular (and now obsolete) execution models, and do not operate on (unannotated) source code.

The complexity of other semantic models of \scoop led to scalability issues when attempting to leverage existing analysis and verification tools. In~\cite{Brooke-PJ07a}, \scoop programs were hand-translated to models in the process algebra \textsc{Csp} to perform, e.g.~deadlock analysis; but the leading \textsc{Csp} tools at the time could not cope with these models and a new tool was purpose-built (but no longer available/maintained today). In a recent deadlock detection benchmark on the \req execution model formalised in \maude~\cite{Morandi-SNM13a}, the tool was not able to give verification results in reasonable time (\ie less than one day) even for simple programs like dining philosophers\footnote{From personal communication with the researchers behind this benchmark.}; our benchmarks compare quite favourably to this. Note that since our work focuses more on semantic modelling and comparisons than it does on the underlying model checking algorithms, we did not yet evaluate the generic bounded model checking algorithms for temporal logic properties implemented in \groove and accessible for \scoopgts models.

\section{Conclusion and Future Work}\label{sec:conclusion}
We proposed and constructed a semantic workbench for a concurrent asynchronous programming language via the following, general work flow:
\begin{inparaenum}[(i)]
  \item derive a \gts-based semantic model from existing semi-formal documentation of execution models;
  \item continuously compare the model by simulation runs against the actual implementations;
  \item exploit semantic paramaterisation to derive a versatile model;
  \item if possible, conduct expert interviews to ascertain the model's faithfulness;
  \item apply existing generic model checking techniques for \gts to implement analyses against the different execution models;
  \item implement different analyses on top of this model.
\end{inparaenum}
This workflow resulted in the formalisation \scoopgts, which covered the two principal execution models of \scoop, and allowed us to formally, automatically, and practically compare the executions of programs with respect to both. With the conducted expert interviews, and the results of applying our model to check the consistency of the semantics across a small but representative collection of \scoop programs in reasonable time, we were reassured of our choice of \gts as an underlying formalism: theoretically sound, yet diagrammatically accessible for software engineers, and able to scale to the sizes of programs we need for semantic comparisons.

We are currently working on extending \scoopgts to cover some more advanced and esoteric features of \scoop (including distributed exception handling) and to enlarge the benchmark set, with the eventual aim of producing a conformance test suite for \scoop-like languages. As noted in~\cite{Zambon-Rensink14a}, the shape of the rules and the control programs have a big influence on the running times of \groove. We are currently working on refactoring \scoopgts for better performance (relative to benchmarking on the conformance test suite).

A more general line of research focuses on the shape of the \scoopgraphs contained in the reachable state space of \scoopgts. Insights here would help us to devise better abstraction techniques (along the lines of~\cite{BackesR15}) with which we could implement better verification algorithms, and visualise the influence of different semantic parameters on \scoopgraphs. Generalising \scoopgts to cover other actor-based concurrency languages would also extend this result towards differences between the semantics of programming language families expressed as \scoopgraph properties.\\


{
\noindent\emph{Acknowledgements.} We thank our interviewees from the \scoop development and research team  for the many helpful and insightful discussions. We are also deeply grateful for the work of the \groove developers that we leverage in this paper, and especially for their \groove-y feedback and support. The underlying research was partially funded by ERC Grant CME \#291389.
}

\bibliographystyle{splncs03}

\bibliography{references}

\end{document}